\documentclass[pdflatex,sn-mathphys-num]{sn-jnl}

\usepackage{graphicx}%
\usepackage{multirow}%
\usepackage{amsmath,amssymb,amsfonts}%
\usepackage{amsthm}%
\usepackage{mathrsfs}%
\usepackage[title]{appendix}%
\usepackage{xcolor}%
\usepackage{textcomp}%
\usepackage{manyfoot}%
\usepackage{booktabs}%
\usepackage{algorithm}%
\usepackage{bm}%
\usepackage{algorithmicx}%
\usepackage{algpseudocode}%
\usepackage{listings}%
\usepackage{lmodern}
\usepackage{anyfontsize}

\newcommand{\sub}[1]{\textcolor{black}{#1}}
\definecolor{sub}{rgb}{0, 0, 0}

\def\argmin{\mathop{\mathrm{arg\,min}}} 

\def\lim{\mathop{\mathrm{lim}}} 

\newcommand{\norm}[1]{\left\lVert#1\right\rVert}

\def\ebm{{\bm{e}}}

\def\xbm{{\bm{x}}}

\def\ybm{{\bm{y}}}

\def\varphibm{{\bm{\varphi}}}

\def\thetabm{{\bm{\theta}}}

\def\Abm{{\bm{A}}}
\def\Hbm{{\bm{H}}}
\def\Dbm{{\bm{D}}}
\def\Fbm{{\bm{F}}}
\def\Sbm{{\bm{S}}}
\def\Pbm{{\bm{P}}}

\def\Ibm{{\bm{I}}}

\def\xbmast{{\bm{x}^\ast}}

\def\xbmhat{{\widehat{\bm{x}}}}

\def\Dsf{{\mathrm{D}}}

\def\Hsf{{\mathrm{H}}}

\def\Rsf{{\mathrm{R}}}

\def\Hsf{{\mathrm{H}}}

\def\C{\mathbb{C}}
\def\R{\mathbb{R}}

\def\Bcal{{\mathcal{B}}}

\usepackage{mathtools}

\usepackage{booktabs}
\usepackage{multirow}       
\usepackage{tabularx}       
\usepackage{hhline}
\usepackage{arydshln}		
\usepackage{xcolor}
\def\Dsf{{\mathsf{D}}}

\def\Rsf{{\mathsf{R}}}

\def\Dsf{{\mathsf{D}}}

\def\Hsf{{\mathsf{H}}}

\def\Hsf{{\mathsf{H}}}

\def\thetabm{{\bm{\theta }}}

\def\Rbf{{\mathbf{R}}}

\def\fbm{{\bm{f}}}

\def\Esf{{\mathsf{E}}}

\def\Xbf{{\mathbf{X}}}

\def\Mbm{{\bm{M}}}

\def\df{g}  
\def\reg{r}
\def\pbmhat{{\bm{\hat{p}}}}

\definecolor{xj}{rgb}{0, 0, 0}
\definecolor{yd}{rgb}{0, 1, 0}
\definecolor{bek}{rgb}{0, 0, 1}

\usepackage{scalerel}
\usepackage{tikz}
\usetikzlibrary{svg.path}

\definecolor{orcidlogocol}{HTML}{A6CE39}
\tikzset{
	orcidlogo/.pic={
		\fill[orcidlogocol] svg{M256,128c0,70.7-57.3,128-128,128C57.3,256,0,198.7,0,128C0,57.3,57.3,0,128,0C198.7,0,256,57.3,256,128z};
		\fill[white] svg{M86.3,186.2H70.9V79.1h15.4v48.4V186.2z}
		svg{M108.9,79.1h41.6c39.6,0,57,28.3,57,53.6c0,27.5-21.5,53.6-56.8,53.6h-41.8V79.1z M124.3,172.4h24.5c34.9,0,42.9-26.5,42.9-39.7c0-21.5-13.7-39.7-43.7-39.7h-23.7V172.4z}
		svg{M88.7,56.8c0,5.5-4.5,10.1-10.1,10.1c-5.6,0-10.1-4.6-10.1-10.1c0-5.6,4.5-10.1,10.1-10.1C84.2,46.7,88.7,51.3,88.7,56.8z};
	}
}

\newcommand\orcidicon[1]{\href{https://orcid.org/#1}{\mbox{\scalerel*{
				\begin{tikzpicture}[yscale=-1,transform shape]
					\pic{orcidlogo};
				\end{tikzpicture}
			}{|}}}}

\usepackage{hyperref} 

\begin{document}
	
	\title[Article Title]{CoRRECT: A Deep Unfolding Framework for Motion-Corrected Quantitative R2* Mapping}
	
	\author[1]{Xiaojian Xu \orcidicon{0000-0002-5264-8963}}\email{xiaojianxu@wustl.edu}
	\author[1]{Weijie Gan \orcidicon{0000-0003-3604-784X}}\email{weijie.gan@wustl.edu}
	\author[2]{Satya~V.V.N.~Kothapalli \orcidicon{0000-0003-0129-3442}}\email{kothapalli.v.v@wustl.edu}
	\author[2]{Dmitriy~A.~Yablonskiy \orcidicon{0000-0003-2898-8797}}\email{yablonskiyd@wustl.edu}
	\author[1,3]{Ulugbek~S.~Kamilov \orcidicon{0000-0001-6770-3278}}\email{kamilov@wustl.edu}
	
	\affil[1]{\orgdiv{Department of Computer Science and Engineering}, \orgname{Washington University in St.~Louis}, \orgaddress{\city{St.~Louis}, \postcode{63130}, \state{MO}, \country{USA}}}
	
	\affil[2]{\orgdiv{Department of Radiology}, \orgname{Washington University in St.~Louis}, \orgaddress{\city{St.~Louis}, \postcode{63130}, \state{MO}, \country{USA}}}
	
	\affil[3]{\orgdiv{Department of Electrical and Systems Engineering}, \orgname{Washington University in St.~Louis}, \orgaddress{\city{St.~Louis}, \postcode{63130}, \state{MO}, \country{USA}}}


	
	\abstract{Quantitative MRI (qMRI) refers to a class of MRI methods for quantifying the spatial distribution of biological tissue parameters. Traditional qMRI methods usually deal separately with artifacts arising from accelerated data acquisition, involuntary physical motion, and magnetic-field inhomogeneities, leading to suboptimal end-to-end performance. This paper presents CoRRECT, a unified deep unfolding (DU) framework for qMRI consisting of a model-based end-to-end neural network, a method for motion-artifact reduction, and a self-supervised learning scheme. The network is trained to produce R2* maps whose k-space data matches the real data by also accounting for motion and field inhomogeneities. When deployed, CoRRECT only uses the k-space data without any pre-computed parameters for motion or inhomogeneity correction. Our results on experimentally collected multi-Gradient-Recalled Echo (mGRE) MRI data show that CoRRECT recovers motion and inhomogeneity artifact-free R2* maps in highly accelerated acquisition settings. This work opens the door to DU methods that can integrate physical measurement models, biophysical signal models, and learned prior models for high-quality qMRI.}
	
	
	
	
	\maketitle
	
	\section{Introduction}
	\label{sec:introduction}
	
	The recovery of diagnostic-quality images from subsampled k-space measurements is fundamental in accelerated \emph{magnetic resonance imaging (MRI)}~\cite{Lustig.etal2007}. The recovery is often viewed as an \emph{inverse problem}, where the unknown image is reconstructed by combining the MRI forward model and a regularizer~\cite{Danielyan.etal2012, Elad.Aharon2006, Hu.etal2012, Rudin.etal1992}. Currently, the state-of-the-art methods for inverse problems are based on \emph{deep learning (DL)}~\cite{Knoll.etal2020, Lucas.etal2018, McCann.etal2017,ongie2020deep, Wang.etal2020}. Traditional DL methods are based on training \emph{convolutional neural networks (CNNs)} to map the measurements to the desired high-quality image. \emph{Deep model-based architectures (DMBAs)}, such as those based on \emph{plug-and-play priors (PnP)} and \emph{deep unfolding (DU)}, have recently extended traditional DL to neural network architectures that combine the MRI forward models and CNN regularizers~\cite{Zhang.Ghanem2018,  Hauptmann.etal2018, Adler.etal2018, Aggarwal.etal2019, Hosseini.etal2019, Yaman.etal2020, Mukherjee.etal2021}
	
	\emph{Quantitative MRI (qMRI)} refers to a class of techniques for quantifying the spatial distribution of biological tissue microstructural parameters from MRI data~\cite{Hernando.etal2012, Zhao.etal2016,Ulrich.Yablonskiy2016, Zhao.etal2017a, Wen.etal2018a,  xiang2020quantitative, Kothapalli.etal2021, Roberts.etal2021}. \sub{qMRI scans are relatively slow due to their reliance on acquisition sequences that require a large number of k-space samples.} \sub{Recovered quantitative maps often suffer from undesirable imaging artifacts caused by measurement noise, macroscopic $B_0$ magnetic field inhomogeneities, and involuntary physical motion during signal acquisition.} There is consequently a need for qMRI methods that can recover high-quality quantitative parameters from accelerated MRI data contaminated by measurement noise, field inhomogeneities, and motion artifacts.
	
	Despite the rich literature on qMRI, the majority of work in the area has considered separately artifacts due to accelerated data acquisition, involuntary physical motion, and magnetic-field inhomogeneities. In particular, it is common to view qMRI parameter estimation as a post-processing step decoupled from the MRI reconstruction~\cite{Zibetti.etal2020, Gao.etal2021, Wu.etal2020a}. We address this issue by presenting a new unified qMRI framework---called \emph{Co-design of MRI Reconstruction and $R_2^\ast$ Estimation with Correction for Motion (CoRRECT)}---for recovering high-quality quantitative $R_2^*$ maps directly from noisy, subsampled, and artifact-corrupted MRI measurements. \sub{Inspired by the state-of-the-art performance of recent DU methods, CoRRECT is developed as a DU framework consisting of three core components: (a) an end-to-end model-based neural network, (b) a training scheme accounting for motion-artifacts, and (c) a  loss function  for training without ground-truth $R_2^*$ maps.} During training, the weights of the CoRRECT network are updated to produce $R_2^\ast$ maps with simulated motion-corrupted k-space data that matches the real motion-corrupted data to account for object motion and magnetic field inhomogeneities. During testing, CoRRECT requires only the k-space data, without any pre-computed parameters related to motion or inhomogeneity correction, thus significantly simplifying and accelerating end-to-end $R_2^\ast$ mapping. We show on experimentally collected \emph{multi-Gradient-Recalled Echo (mGRE)} data that CoRRECT enables the mapping of motion and inhomogeneity artifact-corrected $R_2^*$ maps in highly accelerated acquisition settings. More broadly, this work shows the potential of DU methods for qMRI that can synergistically integrate multiple types of models, including physical measurement models, biophysical signal models, and learned regularization models.

	\sub{The rest of this paper is structured as follows. Sec.~\ref{sec-material} introduces the background and mathematical formulation of the MRI reconstruction and qMRI estimation, along with a discussion of related work and our contributions. Sec.~\ref{sec-method} presents our proposed approach in detail. Sec.~\ref{sec:exp} outlines our experimental setup, presents comparative results with other algorithms, and provides detailed analysis. Finally, Sec.~\ref{sec:conclusion} summarizes our findings and Sec.~\ref{sec:acknowledgment} includes acknowledgments.}
	
	\begin{figure*}[t]
		\centering 
		\includegraphics[width=.99\textwidth]{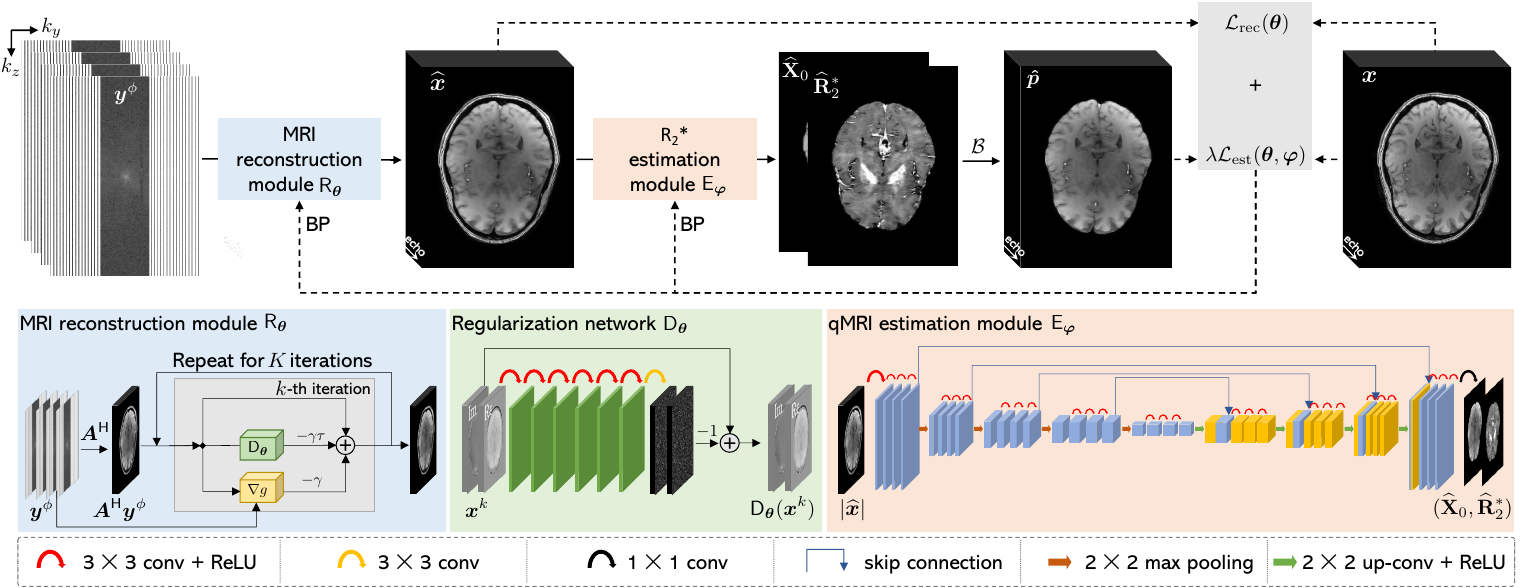}
		\caption{\sub{The overview of the CoRRECT framework for training an end-to-end deep network consisting of two modules: $\Rsf_\thetabm$ for reconstructing mGRE MRI images and $\Esf_\varphibm$ for estimating corresponding $R_2^\ast$ maps. \sub{The network takes as input subsampled, noisy, and motion-corrupted k-space measurements.} $\Rsf_\thetabm$ is implemented as a deep model-based architecture (DMBA) initialized using the zero-filled reconstruction. $\Esf_\varphibm$ is implemented as a customized U-Net architecture mapping the output of $\Rsf_\thetabm$ to the desired $R_2^\ast$ map.The whole network is trained end-to-end using fully-sampled mGRE sequence data without any ground-truth quantitative $R_2^\ast$ maps.}}
		\label{fig:pipeline}
	\end{figure*}
	\section{Background}
	\label{sec-material}
	\subsection{Inverse Problem Formulation}
	
	In MRI, the relationship between the unknown complex-valued image $\xbm \in \C^n$ and its noisy  k-space measurements $\ybm\in\C^m$ is commonly expressed as a linear system
	\begin{equation}
		\label{equ:imaging}
		\ybm = \Abm\xbm + \ebm\ ,
	\end{equation}
	where $\Abm\in\C^{m\times n}$ is the measurement operator and $\ebm\in\C^m$ is the measurement noise, which is often statistically modeled as an additive white Gaussian noise (AWGN). In particular, in multi-coil parallel MRI, the measurement operator $\Abm$ consists of several operators representing the response of each coil~\cite{Fessler2020}
	\begin{equation}
		\label{equ:mcnufft}
		\Abm^i = \Pbm\Fbm\Sbm^i\ ,
	\end{equation}
	where $\Sbm^i$ is the pixel-wise sensitivity map of the $i$th coil, $\Fbm$ is the Fourier transform operator, $\Pbm$ is the k-space sampling operator. When multiple gradient echos are used for qMRI, the sampling pattern $\Pbm$ and the coil sensitivity maps $\{\Sbm^i\}$ are assumed to be fixed for all echo times. We say that the MRI acquisition is ``accelerated'', when each coil collects $m < n$ measurements. It is common to formulate the reconstruction in accelerated MRI as a regularized optimization problem
	\begin{equation}
		\label{equ:optimization}
		\xbmast = \argmin_{\xbm\in\C^n}\ f(\xbm) \quad\text{with}\quad f(\xbm) = \df(\xbm) + \reg(\xbm)\ ,
	\end{equation}
	where $\df$ is the data-fidelity term that quantifies consistency with the measured data $\ybm$ and $\reg$ is a regularizer that enforces a prior knowledge on the unknown image $\xbm$. For example, two widely-used data-fidelity and regularization terms in accelerated MRI are the least-squares and total variation (TV)
	\begin{equation}
		\label{equ:optimization_tv}
		\df(\xbm) = \frac{1}{2}\norm{\ybm-\Abm\xbm}^2_2\quad \mathrm{and}\quad \reg(\xbm)=\tau\norm{\Dbm\xbm}_1\ ,
	\end{equation}
	where $\tau > 0$ controls the regularization strength and $\Dbm$ is the discrete gradient operator~\cite{Rudin.etal1992}.
	
	\subsection{Image Reconstruction using Deep Learning}
	\label{sec:dl-reconstruction}
	
	In the past decade, DL has gained great popularity for solving MRI inverse problems due to its excellent performance (see reviews in~\cite{Lundervold.Lundervold2019, Knoll.etal2020, Zeng.etal2021}). A widely-used supervised DL approach is based on training an image reconstruction CNN $\Rsf_\thetabm$ by mapping a corrupted image $\Abm^\dagger\ybm$ to its clean target $\xbm$, where $\Abm^\dagger$ is an operator that maps the measurements back to the image domain. \sub{The training is formulated as an optimization problem over a training set consisting of desired ground-truth images $\{\xbm_j\}_{j=1}^J$ and their noisy subsampled measurements $\{\ybm_j\}_{j=1}^J$}
	\begin{equation}
		\thetabm^\ast = \argmin_{\thetabm} \sum_{j = 1}^J \mathcal{L}(\Rsf_\thetabm(\Abm^\dagger_j\ybm_j), \xbm_j)\ ,
	\end{equation}
	where $\mathcal{L}$ denotes the loss function that measures the discrepancy between the predictions of the CNN and the ground-truth. Popular choices for the CNN include U-Net~\cite{Ronneberger.etal2015} and for the loss function the $\ell_1$ and $\ell_2$ norms. For example, prior work on DL for accelerated MRI has considered training the CNN by mapping the zero-filled images to the corresponding fully-sampled ground-truth images~\cite{ Wang2016.etal, Han.etal2017, Schlemper.etal2018}.

	PnP~\cite{Venkatakrishnan.etal2013, Sreehari.etal2016} is a widely-used framework that extend the traditional DL by enabling the integration of the physical measurement models and powerful CNN denoisers as image priors to provide state-of-the-art reconstruction algorithms (see recent reviews of PnP in~\cite{Ahmad.etal2020, Kamilov.etal2022}). For example, a well-known PnP method \emph{regularization by denoising (RED)}~\cite{Romano.etal2017} can be expressed as
	\begin{equation}
		\label{Eq:REDIteration}
		\xbm^k \leftarrow  \xbm^{k-1} - \gamma \left(\nabla \df(\xbm^{k-1}) + \tau(\xbm^{k-1} - \Dsf_\thetabm(\xbm^{k-1}))\right)\;,
	\end{equation}
	where $\nabla g$ is the gradient of the data-fidelity term in Eq.~\eqref{equ:optimization}, $\Dsf_\thetabm$ is the CNN denoiser parameterized by weights $\thetabm$, 
	and $\gamma, \tau > 0$ are the step size and  the regularization parameters, respectively. The iterates of Eq.~\eqref{Eq:REDIteration} seek an equilibrium between the physical measurement model and learned prior model. Remarkably, this heuristic of using CNNs not necessarily associated with any $\reg$ within an iterative algorithm exhibited great empirical success~\cite{Sreehari.etal2016,Zhang.etal2017a, Zhang.etal2019, Ahmad.etal2020} and spurred a great deal of theoretical work on PnP~\cite{chan2016plug, Ryu.etal2019, Xu.etal2020,Sun.etal2021a}. 
	
	DU (also known as \emph{deep unrolling} and \emph{algorithm unrolling}) is another widely-used DL paradigm that was widely adopted in MRI due to its ability to provide a systematic connection between iterative algorithms and deep neural network architectures~\cite{Yang.etal2016, Zhang.Ghanem2018,  Hauptmann.etal2018, Adler.etal2018, Aggarwal.etal2019, Hosseini.etal2019, Yaman.etal2020, Mukherjee.etal2021}. PnP algorithms can be naturally turned into DU architectures by truncating the PnP algorithm to a fixed number of iterations and training the corresponding architecture end-to-end in a supervised fashion. By training the CNN  $\Dsf_\thetabm$ jointly with the measurement model, DU leads to an end-to-end network optimized for a given inverse problem.

	\subsection{Deep qMRI Map Estimations }
	\label{sec-background-est}
	qMRI maps are traditionally obtained by fitting a biophysical model to MRI images in a voxel-by-voxel fashion. Traditional fitting methods are time consuming and are sensitive to the artifacts in MRI images (e.g., noise or motion). \sub{Recent work has shown the effectiveness of \emph{deep neural networks (DNNs)} for estimating high-quality qMRI maps (see recent reviews in~\cite{Liu2020, Jung.etal2022, Feng.etal2022, kofler2024machine}).} One conventional application of DL in qMRI seeks to train a DNN to learn a direct mapping of qMRI maps from the MR images in a supervised fashion. The training can be guided by minimizing the loss between the outputs of the DNN and the qMRI maps estimated from the MR images using standard fitting methods. This end-to-end mapping strategy has been investigated in several qMRI applications, including $T_2$~\cite{Cai.etal2018}, high quality susceptibility  mapping (QSM)~\cite{Yoon.etal2018, Bollmann.etal2019}, $T_1$ and $T_{1\rho}$~\cite{li2020ultra}, $R2t^\ast$ and $R2'$~\cite{Kahali2021.09.10.459810}. It has also been applied to help magnetic resonance fingerprinting (MRF)~\cite{Panda.etal2017} with a better and more efficient generation of qMRI maps such as $T_1$ and $T_2$~\cite{Cohen.etal2018, Fang.etal2019}.  \sub{The work in~\cite{Torop.etal2020, Jeelani.etal2020, Xu.etal2022} have explored the potential of self-supervised learning for training qMRI estimation networks directly on MRI images using biophysical models without ground truth qMRI maps. Particularly, it has been shown that the self-supervised loss is equivalent to the supervised loss for the $R_2^\ast$ mapping in~\cite{Torop.etal2020}. Recent work has also focused on DL-based image reconstruction methods for improving qMRI estimation~\cite{Zibetti.etal2020, Gao.etal2021, Wu.etal2020a}. }
	
	\textcolor{xj}{When the measurement operator $\Abm$ is available, it can be combined with biophysical models to enforce data consistency relative to the subsampled measurements, leading to a model-based qMRI mapping. For example, the work in~\cite{Liu.etal2019} developed a CNN to directly convert a series of undersampled MR images straight into qMRI parameter maps using supervised training, while the end-to-end CNN mapping also serves as a data consistency component by fitting to the subsampled k-space measurements. Building on the foundation established in~\cite{Liu.etal2019}, the study in~\cite{Liu.etal2020b} further introduces an additional adversarial loss highlighting that estimated maps resemble the same tissue features and image sharpness as the reference maps from fully-sampled data for improved detail preservation.} \sub{The self-supervised method established in~\cite{Liu.etal2021b} follows a similar idea of using a CNN to directly map the subsampled MR images to high-quality qMRI maps.} However, instead of using reference qMRI maps or fully-sampled data, the network in~\cite{Liu.etal2021b} is trained by fitting the output to the subsampled k-space data using qMRI biophysical model and MR imaging model where generic sparsity constraints, such as TV, can be additionally included to regularize the performance. \sub{The work in~\cite{zimmermann2024pinqi} introduces an approach that alternates between MRI image reconstruction and qMRI parameter estimation, both utilizing CNN-based regularization, enabling end-to-end training.} Despite the active research in the area, we are not aware of any prior work that has developed a DMBA combining the DL-based MRI reconstruction and qMRI estimation into a single imaging pipeline that can address in a unified fashion k-space subsampling, measurement noise, object motion, and magnetic-field inhomogeneities.

	\subsection{Our Contribution}
	\label{sec-background-cont}
	
	This work contributes to the rapidly evolving area of qMRI parameter estimation using DL. We introduce a new framework, called CoRRECT, for the mapping of $R_2^\ast$ maps directly from artifact-corrupted k-space measurements. CoRRECT addresses in a unified fashion several common sources of image artifacts, including those due to k-space subsampling, measurement noise, object motion, and magnetic field inhomogeneities, which has never been done before. CoRRECT can be viewed as a DMBA that enables a principled integration of several types of mathematical models for qMRI, including MRI forward model, object motion model, mGRE biophysical model, and a learned prior model characterized by a CNN.  \textcolor{xj}{Although learning-based parameter mapping are not new ideas,  there is no work that has investigated the unified DU-based MR image reconstruction and qMRI parameter mapping along with motion correction. \sub{By incorporating motion during the training, our work tackles such practical challenges and establishes the effectiveness of the learning-based method in handling both motion-corrected MR image reconstruction and qMRI parameter mapping in an end-to-end fashion.}  We summarize the key features and contributions of this work as follows:}
	\begin{itemize}
		\item \textcolor{xj}{CoRRECT realizes, to the best of our knowledge,  the first end-to-end estimation for high-quality $R_2^\ast$ maps directly form subsampled, noisy, and motion-corrupted k-space measurements.} 
		
		\item \textcolor{xj}{CoRRECT allows the simultaneous MRI reconstruction and $R_2^\ast$ estimation, and meanwhile the motion correction for both. Our architecture consists of a DU-based MRI reconstruction module and a CNN-based $R_2^\ast$ estimation module that can easily be modified for the estimation of different qMRI parameters other than $R_2^\ast$.}
		
		\item \textcolor{xj}{We propose a mGRE-guided loss function that enables the training of CoRRECT without the access to the ground-truth $R_2^\ast$ maps. This capability is achieved by adopting recent ideas from self-supervised deep learning~\cite{Xu.etal2022, Torop.etal2020}.} 
		
		\item \textcolor{xj}{We extensively validate CoRRECT on both simulated and experimental data. Our results show that CoRRECT both quantitatively and qualitatively outperforms the popular baseline methods and leads to significant quality improvements in both MRI reconstruction and $R_2^\ast$ estimation.}
	\end{itemize}

	\section{Proposed Method}
	\label{sec-method} 
	The CoRRECT framework introduced in this section consists of several modules that enable end-to-end training of a model-based architecture for the mapping of $R_2^\ast$ maps.
	
	\subsection{CoRRECT Biophysical Model}
	\label{sec:biophysicalmodel}
	
	CoRRECT is based on the \emph{multi-Gradient-Recalled Echo (mGRE)} sequences for $R_2^\ast$ mapping. mGRE is a wiedely-used sequence for producing quantitative maps related to biological tissue microstructure in health and disease~\cite{Hernando.etal2012, Zhao.etal2016,Ulrich.Yablonskiy2016, Zhao.etal2017a, Wen.etal2018a,  xiang2020quantitative, Kothapalli.etal2021, Roberts.etal2021}. For $R_2^\ast$ mapping, each reconstructed mGRE voxel can be interpreted using the following \emph{biophysical model}~\cite{Yablonskiy1998}
	\begin{equation}
		\label{Eq:Model}
		x(t) = X_0 \cdot \exp(-R_2^\ast \cdot t - i {\omega}  t) \cdot F(t),
	\end{equation}
	where $t$ denotes the gradient echo time, $X_0 = x(0)$ is the signal intensity at $t = 0$, and $\omega$ is a local frequency of the MRI signal. \sub{Note that Eq.~\eqref{Eq:Model} is a voxel-based model, where $x(t)$ represents a single voxel from the unknown $N$-echo mGRE image $\xbm$ in Eq.~\eqref{equ:imaging}, evaluated at a specific echo time $t$.} The complex  valued function $F(t)$ in Eq.~\eqref{Eq:Model} models the effect of macroscopic magnetic field inhomogeneities on the mGRE signal. The failure to account for such inhomogeneities is known to bias and corrupt the recovered $R_2^\ast$ maps. The function $F(t)$ is traditionally computed using the \emph{voxel spread function (VSF)} approach~\cite{Yablonskiy.etal2013}, based on evaluating the effects of macroscopic magnetic field inhomogeneities (background gradients) on formation of the complex-valued mGRE signal. The $R_2^\ast$ maps, $\omega$ maps, and $X_0$  can be jointly estimated from 3D mGRE data acquired at different echo times $t$ by fitting Eq.~\eqref{Eq:Model} with pre-calculated $F(t)$ on a voxel-by-voxel basis using \emph{non-linear least squares (NLLS)}~\cite{Yablonskiy.etal2013}. However, NLLS fitting is time-consuming and sensitive to the  artifacts in MRI images. In this work, we use CoRRECT to enable the learning-based high-quality mGRE reconstruction and $R_2^\ast$ estimation directly from the artifact-corrupted k-space measurements.
	\sub{For simplicity, we refer to Eq.~\eqref{Eq:Model} as the biophysical model $\Bcal$ in the following discussion.}

	\color{sub}
	\subsection{CoRRECT Architecture}
	CoRRECT considers a  motion-corrupted version of Eq.~\eqref{equ:imaging}, where the motion-corrupted measurements $\ybm^\phi$ are simulated by replacing regions of the motion-free k-space data of $\xbm$ with its moved version $\phi(\xbm)$ as
	\begin{equation}
		\label{equ:pro-fwd-with-motion}
		\ybm^\phi = (\Ibm - \sum_{l=1}^{L} \Hbm_{l}) \Abm\xbm + \sum_{l=1}^{L}\Hbm_{l}\Abm\phi(\xbm) + \ebm\;.
	\end{equation}
	Here, $\Ibm \in \R^{m \times m}$ is an identity matrix,  $L \geq 1$ denotes the total number of all motion events, and the function $\phi$ denotes the effect of an unknown motion of the object during scanning. The diagonal matrix $\Hbm_{l} \in \{0, 1\}^{m \times m}$ denotes a binary map where its diagonal elements have values 1 in locations corrupted by the $l$-th motion event, simulating the beginning and duration of this motion. Our imaging pipeline produces both the motion-corrected mGRE image and the corresponding $R_2^\ast$ map by training a DMBA on a set of ground-truth mGRE images $\{\xbm_j\}_{j=1}^J$ and their noisy, motion-corrupted and subsampled measurements $\{\ybm_j^\phi\}_{j=1}^J$ given the measurement operators $\{\Abm_j\}_{j=1}^J$ for each acquisition. Fig.~\ref{fig:pipeline} summarizes the  CoRRECT framework by omitting the sample index $j$ for simplicity. 
	
	\subsubsection{Motion Simulation}
	Building upon the motion corruption model in Eq.~\eqref{equ:pro-fwd-with-motion}, we obtain a training dataset consisting of pairs of ground-truth mGRE images $\xbm_j$ and their subsampled, noisy, and motion-corrupted measurements $\ybm^\phi_j$ (see details in Sec. ~\ref{sec:data_simu}) as
	\begin{equation}
		\label{eq:motion}
		\ybm^\phi_j = (\Ibm - \sum_{l=1}^{L} \Hbm_{jl}) \Abm_j\xbm_j + \sum_{l=1}^{L}\Hbm_{jl}\Abm_j\phi(\xbm_j) + \ebm_j\;.
	\end{equation}
	Our model is trained by using $\xbm_j$ as a label for the motion-corrupted input measurement $\ybm^\phi_j$, enabling motion-corrected mapping of $R_2^\ast$.
	\color{black}

	\subsubsection{mGRE Reconstruction Module}
	\label{sec-method-rec}
	The mGRE reconstruction module $\Rsf_\thetabm$ seeks to produce high-quality $N$-echo mGRE image $\xbmhat_j$ given the subsampled, noisy, and motion-corrupted k-space measurements $\ybm^\phi_j$ and the measurement operator $\Abm_j$
	\begin{equation}
		\label{eq:mri_output}
		\xbmhat_j = \Rsf_\thetabm(\ybm^\phi_j;\Abm_j)\;.
	\end{equation}
	This module is a $K$-layer architecture consisting of two types of sub-modules: (a) data-consistency sub-module for ensuring that predicted mGRE images match the k-space measurements; (b) regularization sub-module consisting of a CNN $\Dsf_\thetabm$ with trainable parameters $\thetabm  \in \R^p$. The data-consistency sub-module is implemented as the gradient-step of the least-squares penalty Eq.~\eqref{equ:optimization_tv}
	\begin{align}
		\label{Eq:grad_g}
		\nabla \df(\xbm_j) = \Abm_j^\Hsf (\Abm_j \xbm_j - \ybm_j^\phi)\ ,
	\end{align}
	where $\Abm_j^\Hsf$ denotes the hermitian transpose of the measurement operator $\Abm_j$. As shown in Fig.~\ref{fig:pipeline}, the reconstruction module is initialized as $\Abm^\Hsf_j\ybm^\phi_j$. In our implementation, we set $K = 8$ and use a customized DnCNN~\cite{Zhang.etal2017} for the implementation of the regularization network $\Dsf_\thetabm$. \textcolor{xj}{Our customized $\Dsf_\thetabm$ consists of 7 layers, where each of the first 6 layers is a convolution  (conv) layer followed by rectified linear unit (ReLU), and the last is a convolution layer. } \sub{Kernel sizes of all convolutions are set to 3, strides to 1, and number of filters to 64.} $\Dsf_\thetabm$ is implemented using the strategy of residue learning, where its outputs are the artifacts in the inputs, and the clean predictions are obtained by subtracting those artifacts from the inputs. The weights of $\Dsf_\thetabm$ are shared across all $K$ steps for memory efficiency. To enable the reconstruction for complex mGRE data,  the  input  of $\Dsf_\thetabm$ are split to 2 channels that consist of the real (denoted as $\operatorname{Re}$) and imaginary (denoted as $\operatorname{Im}$) parts.

	\subsubsection{$R_2^\ast$ Estimation Module}
	\label{sec-method-est}
	The  mGRE reconstruction module $\Rsf_\thetabm$ is followed by a $R_2^\ast$ estimation module $\Esf_\varphibm$ that produces $R_2^\ast$ from mGRE images. We implement $\Esf_\varphibm$ using a CNN customized from U-Net~\cite{Ronneberger.etal2015} with trainable parameters $\varphibm \in \R^q$.  Our customized U-Net consists of five encoder blocks, four decoder blocks with skip connections, and an output block. For each block in encoder and decoder blocks, it consists of convolutions followed by ReLU. Kernel sizes of all convolutions are set to 3 and strides to 1. The number of filters are set to 64, 128, 256, 512, and 1024 in each encoder block and to 512, 256, 128, and 64 in each decoder block sequentially from inputs to outputs. The network $\Esf_\varphibm$ takes the magnitude of the reconstructed $N$-echo mGRE image $\xbmhat_j$ from $\Rsf_\thetabm$ as the input and produces the qMRI maps $(\widehat{\Xbf}_0, \widehat{\Rbf}_2^\ast)_j$ as the output
	\begin{equation}
		\label{eq:qmri_output}
		(\widehat{\Xbf}_0, \widehat{\Rbf}_2^\ast)_j= \Esf_\varphibm(|\xbmhat_j|) \;,
	\end{equation}
	where $|\cdot|$ returns the magnitude of its input vector, and $\widehat{\Xbf}_0 \in \R^n, \widehat{\Rbf}_2^\ast \in \R^n$ denote the vectorized $X_0$ and $R_2^\ast$ outputs from the estimation module, respectively. 
	

	\subsection{CoRRECT Training}
	\textcolor{xj}{We adopt a mGRE-guided loss to train the CoRRECT architecture end-to-end, where only the mGRE images are used for training without any ground-truth $(X_0, R_2^\ast)$ maps.} Consider the intermediate mGRE output $\xbmhat_j$ produced by $\Rsf_\thetabm$ in Eq.~\eqref{eq:mri_output} and the corresponding quantitative map $(\widehat{\Xbf}_0, \widehat{\Rbf}_2^\ast)_j$ produced by $\Esf_\varphibm$ in Eq.~\eqref{eq:qmri_output}. The training is implemented by minimizing two loss functions: the mGRE reconstruction loss $\mathcal{L}_\mathrm{rec}(\thetabm)$ and the $R_2^\ast$ estimation loss $\mathcal{L}_\mathrm{est}(\thetabm, \varphibm)$. Given data $\xbm_j$ and $\ybm^\phi_j$, the mGRE reconstruction loss $\mathcal{L}_\mathrm{rec}(\thetabm)_j$ measures the difference between the reconstructed mGRE image $\xbmhat_j$ and the ground truth mGRE image $\xbm_j$ as
	\begin{equation}
		\mathcal{L}_{\mathrm{rec}}(\thetabm)_j
		= \mathcal{L}(\xbmhat_j, \xbm_j)\ .
	\end{equation} 
	The $R_2^\ast$ estimation loss $\mathcal{L}_{\mathrm{est}}(\thetabm, \varphibm)_j$ enforces consistency of the produced mGRE images from the estimated $(X_0, R_2^\ast)$ maps to the ground-truth mGRE images. The estimation loss uses the biophysical model ${\pbmhat_j 
		= \Bcal((\widehat{\Xbf}_0, \widehat{\Rbf}_2^\ast)_j ; \fbm_j)}$ in Eq.~\eqref{Eq:Model} to relate the mGRE images and the quantitative $R_2^\ast$ maps
	\begin{equation}
		\label{eq:loss-est}
		\mathcal{L}_{\mathrm{est}}(\thetabm, \varphibm)_j
		=\mathcal{L}(|\Mbm_j\pbmhat_j|, |\Mbm_j\xbm_j|),
	\end{equation}
	where $\fbm_j \in \C^n$ denotes the vectorized $F(t)$ function pre-computed using the VSF approach~\cite{Yablonskiy.etal2013} from ground-truth mGRE data $\xbm_j$ to compensate for the effect of macroscopic magnetic field inhomogeneities, and $\Mbm_j$ denotes the voxel-wise region extraction mask (REM) where the biophysical model applies. Note that $\Mbm_j$ and $\{\fbm_j\}$ are only needed during training for evaluating the $R_2^\ast$ estimation loss, but not during the inference. Given losses $\mathcal{L}_\mathrm{rec}(\thetabm)_j$ and $\mathcal{L}_\mathrm{est}(\thetabm, \varphibm)_j$, CoRRECT training seeks to minimize their combination over a training set consisting of $J$ samples
	\begin{equation}
		\thetabm^\ast, \varphibm^\ast = \argmin_{\thetabm, \varphibm} \sum_{j = 1}^J  \left\{ \mathcal{L}_\mathrm{rec}(\thetabm)_j + \lambda \mathcal{L}_\mathrm{est}(\thetabm, \varphibm)_j\right\},
	\end{equation}
	where $\lambda >0$ is a weight parameter. The learned parameters $\thetabm^\ast$ and $\varphibm^\ast$ can be computed by using gradient-based minimization algorithms such as SGD or Adam.
	
	\textcolor{xj}{It is worth highlighting that CoRRECT does not need ground-truth quantitative $R_2^\ast$ maps during training.} Instead, it is trained using only mGRE images and our knowledge of the biophysical model $\Bcal$ connecting the mGRE signal with $R_2^\ast$. The biophysical model accounts for magnetic field inhomogeneities by using $F(t)$, which enables the estimation module $\Esf_\varphibm$ to compensate for macroscopic magnetic field inhomogeneities, thus producing motion-artifact-corrected and $B0$-inhomogeneity-corrected $R_2^\ast$ maps.  Note that $F(t)$ is only required during training not inference, enabling fast end-to-end estimation of quantitative maps. As corroborated by our empirical results, the joint training of $\Rsf_\thetabm$ and $\Esf_\varphibm$ within CoRRECT leads to better overall performance.

	\color{sub}
	\subsection{Mathematical Analysis of CoRRECT}
	\label{sec: mathematical}
	As a downstream task dependent on reconstructed MR images, qMRI parameter estimation is traditionally approached in two independent steps: (a) reconstructing artifact-free MR images and (b) estimating quantitative parameter maps. This separation limits the reconstruction process from leveraging task-specific prior knowledge that is crucial for accurate qMRI parameter estimation. To address this issue, CoRRECT integrates a reconstruction network $R_\thetabm$ and a qMRI estimation network $E_\varphibm$ into a unified training pipeline, enabling direct mapping from motion-corrupted raw $k$-space data to high-quality qMRI maps. Mathematically, this joint training can be interpreted as the learning of a task-aware reconstruction network $R_\thetabm$ that incorporates the MRI forward model $\Abm$, the motion-corrupted measurements $\ybm^\phi$, and the qMRI estimation network $E_\varphibm$:
	\begin{equation}
		\label{eq:joint-regularizer}
		E_\varphibm \left(R_\thetabm(\ybm^\phi; \Abm)\right)  \text{ where } R_\thetabm(\ybm^\phi; \Abm) \in \mathop{\mathsf{argmin}}_{\xbm} \left\{\df(\xbm) + r_\thetabm(\xbm)\right\}\;.
	\end{equation}
	Here, we use $r_\thetabm$ to denote an implicit regularizer, which is approximated by the CNN-based regularizer within $R_\thetabm$. Unlike traditional disjoint approaches where $r_\thetabm$ is optimized solely for image reconstruction, our joint training framework enables the reconstruction process to learn a regularizer $r_\thetabm$ that operates not only in the image space of $\xbm$ but also aligns with the quantitative parameter space. This alignment can substantially improve the downstream qMRI estimation task. Additionally, CoRRECT also trains $r_\thetabm$ to  compensate for object motions, which further refines $r_\thetabm$ to address  corresponding motion artifacts throughout the reconstruction. In summary, by aligning the reconstruction process with the downstream task, CoRRECT effectively learns an $r_\thetabm$ that simultaneously enhances both high-quality MRI reconstruction and qMRI parameter estimation. Our experimental results in the following section demonstrate that this joint training approach significantly enhances qMRI parameter estimation performance, highlighting the advantages of CoRRECT's integrated design.
	\color{black}

	\section{Experimental Validation}
	\label{sec:exp}
	
	In this section, we present numerical results on simulated and experimentally collected mGRE data showing the ability of CoRRECT to provide high-quality $R_2^\ast$ maps from subsampled, noisy, and motion-corrupted k-space measurements.

	\subsection{Dataset Preparation}
	We used fully-sampled k-space mGRE data of the brain to generate the synthetic subsampled, noisy, and motion-corrupted measurements. These brain data were collected from 15 healthy volunteers using a Siemens 3T Trio MRI scanner and a 32-channel phased-array head coil. Studies were conducted with the approval of the local IRB of Washington University. All volunteers provided informed consent. The data were obtained using a 3D version of the mGRE sequence with $N = 10$ gradient echoes followed by a navigator echo~\cite{Wen.etal2015} used to reduce artifacts induced by physiological fluctuations during the scan. Sequence parameters were flip angle $FA = 30^\circ$, voxel size of $1 \times 1 \times 2$ mm$^3$, first echo time $t_1 = 4$ ms, echo spacing $\Delta t = 4$ ms (monopolar readout), repetition time TR $= 50$ ms.  The dimension of raw measurement for each subject from each coil at a single  echo time was $N^{k_x} \times N^{k_y}  \times N^{k_z} $ with $k_x$ and $k_y$ both being the phase-encoding dimension and $k_z$ being read-out (frequency-encoding) dimension, respectively. \sub{In our data, $N^{k_x} = 72$, $N^{k_y} = 192$, and $N^{k_z} = 256$, resulting in the dimension of $\xbm \in \C^n$ in Eq.~\eqref{equ:imaging} being $n = N^{k_x} \times N^{k_y} \times N^{k_z} \times N = 72 \times 192 \times 256 \times 10$.} Due to limited GPU memory, we converted 3D k-space data into 2D slices by taking a 1D Fourier Transform along the $k_x$ dimension and performed 3D MRI reconstruction and $R_2^\ast$ estimation in a slice-by-slice manner. 
	
	The 15 subjects were split into 10, 2, and 3 for training, validation, and testing, respectively. \sub{For each subject, we extracted the middle 25 to 55 slices (72 in total) of the brain volume that contained the most relevant regions of the brain.} \textcolor{xj}{This produced 3100 images for training, 620 for validation, and 930 for testing.} For each slice, $10$-echo mGRE images of fully-sampled, noise- and motion-free k-space data was used as the ground truth,  corresponds to the target image $\xbm$ in Eq.~\eqref{equ:pro-fwd-with-motion}.  The ground-truth images $\xbm$ deformed with the synthetic motion function $\phi$, measured using the forward operator $\Abm$, and contaminated by AWGN in Eq.~\eqref{equ:pro-fwd-with-motion} to generate artifact-corrupted measurements $\ybm^\phi$ (see Sec.~\ref{sec:data_simu}). The data $\{ \xbm_j, \ybm_j^\phi\}$ of all samples were used for training and quantitative evaluation.  Additional experimental data with clearly visible motion artifacts were used for experimentally evaluating the performance of our network. \sub{Similar to the collection process for motion-free data, the motion-corrupted experimental data were gathered with volunteers instructed to lie still.} However, natural actions such as swallowing, sneezing, or slight head movements inevitably occurred, resulting in realistic motion-corrupted data. The coil sensitivity maps for each slice were estimated from its $1$st echo of fully sampled k-space data using ESPIRiT~\cite{Uecker.etal2014} for both simulated and experimental data.

	\begin{table*}[t]
		\centering
		
		{
			\footnotesize
			\begin{threeparttable}
				\caption{\textcolor{sub}{Quantitative evaluation of CoRRECT on simulated data at different sampling rates. Note that CoRRECT achieves the highest averaged SNR and SSIM compared with all the baseline methods in terms of both mGRE reconstruction and $R_2^\ast$ estimation.}}
				\label{tab:simu}
				\renewcommand\arraystretch{1}
				\setlength{\tabcolsep}{8pt}
				\begin{tabular}{ccccccc} 
					\toprule
					\textit{Images}       & \multicolumn{6}{c}{mGRE} \\ 
					\cmidrule(rl){2-7}
					\textit{Metric}   &\multicolumn{3}{c}{SNR (dB)} & \multicolumn{3}{c}{SSIM}\\
					\cmidrule(rl){2-7}
					\textit{Acceleration rate}   & x2 & x4 & x8 & x2 & x4 & x8  \\
					\cmidrule(rl){2-4}\cmidrule(rl){5-7}
					ZF+NLLS & 16.72 & 14.73 & 14.00 & 0.90 & 0.86 & 0.85  \\
					TV      & 21.46 & 19.88 & 17.05 & 0.81 & 0.8 & 0.77 \\
					RED     & 21.49 & 20.10 & 17.49 & 0.92 & 0.90 & 0.87 \\
					U-Net   & 20.79 & 19.25 & 18.09 & 0.92 & 0.90 & 0.88  \\
					DU      & 21.53 & 20.36 & 19.08 & 0.93 & 0.91 & 0.90 \\
					CoRRECT (Ours) & \textbf{22.12} & \textbf{20.66} & \textbf{19.25} & \textbf{0.93} & \textbf{0.91} & \textbf{0.90}  \\
					

					\bottomrule
				\end{tabular}
				
				\begin{tabular}{ccccccc} 
					\toprule
					\textit{Images}       & \multicolumn{6}{c}{$R_2^\ast$} \\ 
					\cmidrule(rl){2-7}
					\textit{Metric}   &\multicolumn{3}{c}{SNR (dB)} & \multicolumn{3}{c}{SSIM}\\
					\cmidrule(rl){2-7}
					\textit{Acceleration rate}   & x2 & x4 & x8 & x2 & x4 & x8  \\
					\cmidrule(rl){2-4}\cmidrule(rl){5-7}
					ZF+NLLS &6.70 & 6.30 & 6.17 & 0.85 & 0.82 & 0.82   \\
					TV      &12.21 & 11.72 & 10.60 & 0.92 & 0.90 & 0.87 \\
					RED     &12.16 & 11.70 & 10.59 & 0.91 & 0.90 & 0.87 \\
					U-Net   &12.08 & 11.39 & 10.77 & 0.91 & 0.89 & 0.88  \\
					DU      &12.20 & 11.79 & 11.15 & 0.92 & 0.90 & 0.89 \\
					CoRRECT (Ours) & \textbf{12.99} & \textbf{12.33} & \textbf{11.60} & \textbf{0.92} & \textbf{0.90} & \textbf{0.89}  \\
					
					\bottomrule
				\end{tabular}
			\end{threeparttable}
		}
	\end{table*}

	\subsection{Training Data Simulation}
	\label{sec:data_simu}
	
	In order to train our architecture, we need matched pairs of clean mGRE images $\xbm$ and corresponding k-space measurements $\ybm^\phi$. This section discusses our simulation pipeline for generating such training data that accounts for subsampling, noise, and motion artifacts.

	\subsubsection{Accounting for Object Motion}
	
	The motion artifacts in k-space measurements were modeled as a series of physical motions, such as shifts or rotations, that result in the perturbation of blocks of k-space lines. We implemented this process by replacing sections of k-space lines of the ground-truth MR images $\xbm$ with those from their moved versions to synthesize motion artifacts. To generate a wide-range of artifacts, we controlled the number of movements, the duration of each movement, and the amplitude of each movements as random numbers following the configuration in~\cite{Xu.etal2022}. \sub{Specifically, we selected the total number of motions occurring during data acquisition as a random number in the range from 1 to 10. For each motion, we simulated random in-plane shifts within the range of 0 to 15 voxels followed by a combination of random rotations along each axis relative to the center of a  3D mGRE data volume, where each rotation was within the range of 0$^{\circ} $ to 15$^{\circ}$. The time at which each motion occurred and the duration it lasted were randomly generated as well. In particular, all motions were assumed to occur randomly throughout the whole examination process, and each of them was assumed to last for a random duration from about 3 seconds to 30 seconds, which would be equivalent to disturbing about 1 to 10 k-space lines in a single 2D slice. All random numbers mentioned above were uniformly generated in the given range, introducing various levels of motion artifacts to our training and validation dataset. Considering the fact that k-space scanning in the echo direction is much faster than the physical movement, we assumed that all 10-echo images of a data slice suffered from the same motion effects. While the simulation setting above yielded excellent performance in our experimental data, it can be adjusted for different applications.}
	
	\begin{figure*}[t]
		\centering 
		\includegraphics[width=\textwidth]{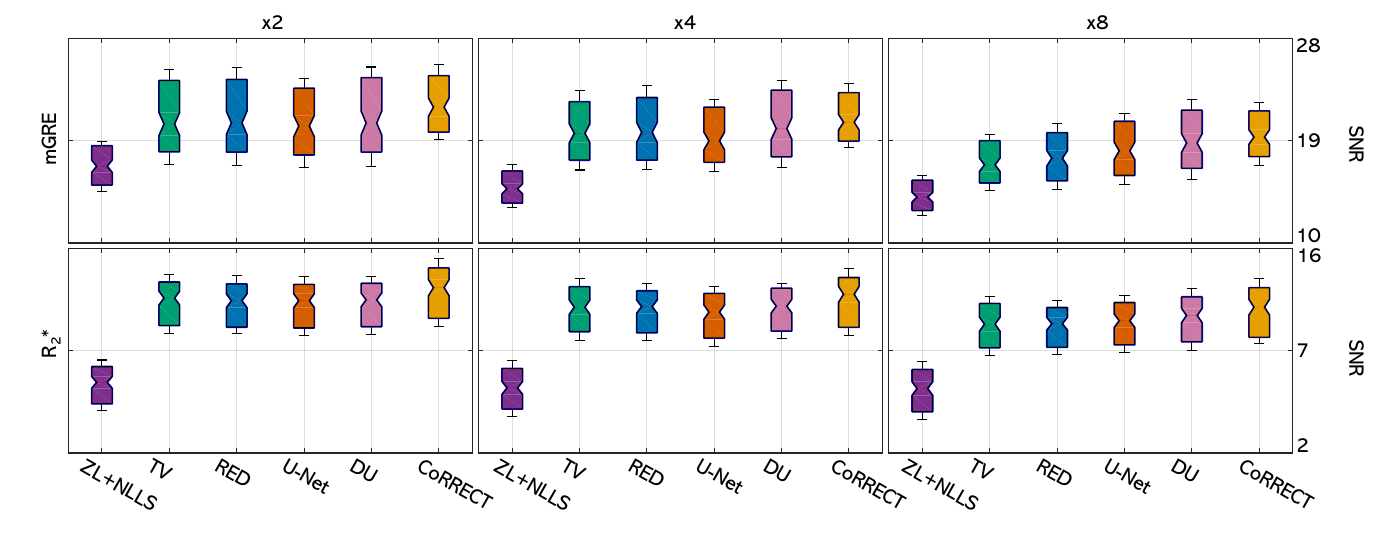}
		\caption{\sub{The statistical box plot of SNR values for different methods obtained on simulated data at different sampling rates. Results highlight the performance of CoRRECT in both mGRE reconstruction and  $R_2^\ast$ estimation against different approaches.}}
		\label{fig:statistic}
	\end{figure*}
	\subsubsection{Accounting for Subsampling and Noise}
	The k-space measurements contaminated by simulated motion were further subsampled and contaminated by AWGN. We used a Cartesian sampling pattern that fully-samples along $k_x$ and $k_z$ dimensions, and subsamples along the $k_y$ dimension. We considered three sampling rates $50\%$,  $25\%$, and $12.5\%$, which we referred to as acceleration rates $\times2$, $\times4$ and $\times8$, respectively. For each rate, we kept the central 60 out of 192 lines along $k_y$.The simulation $\ybm^\phi$ finally included the addition of AWGN corresponding to an input SNR of 40dB.
	
	
	\begin{figure*}[t]
		\centering 
		\includegraphics[width=\textwidth]{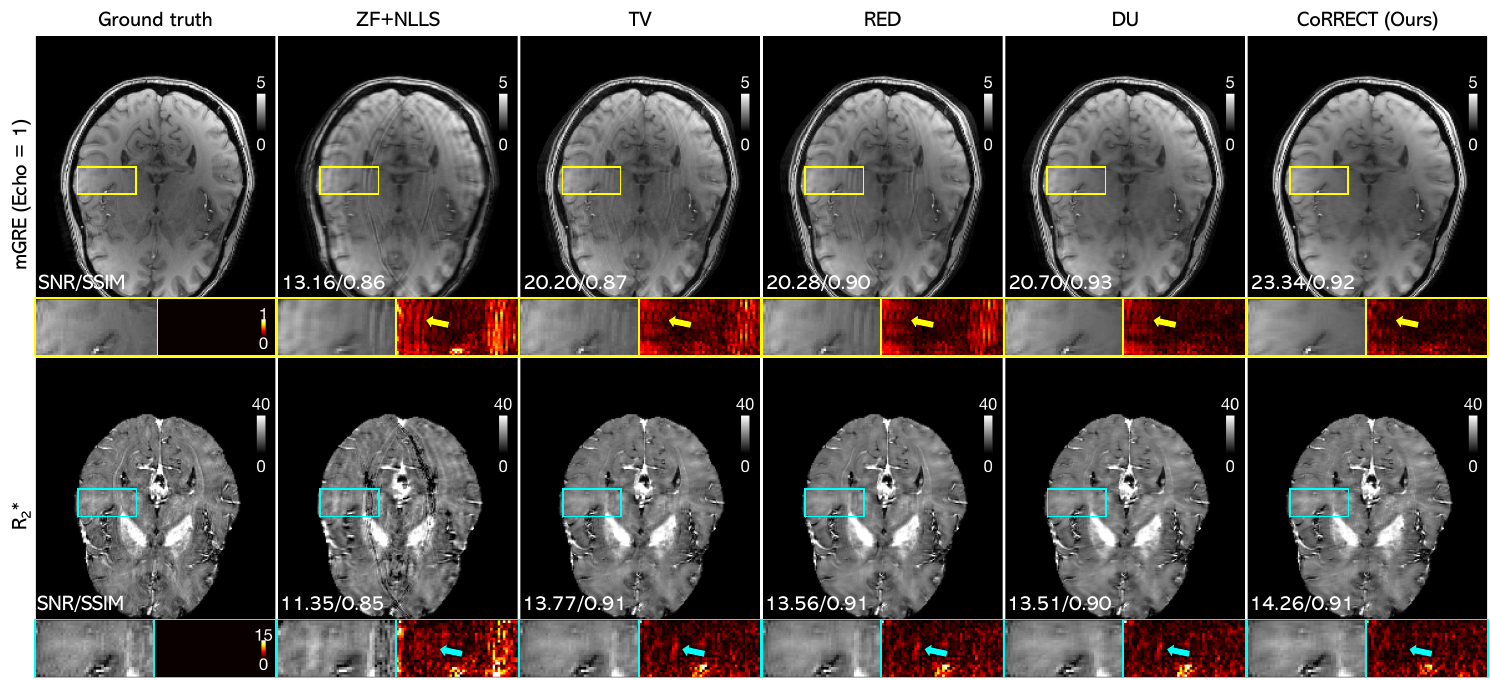}
		\caption{Quantitative and visual evaluation of CoRRECT on simulated data corrupted with synthetic motion, sampled using acceleration factor $\times 4$. The bottom-left corner of each image provides the SNR and SSIM values with respect to the ground-truth.  Arrows in the zoomed-in plots highlight brain regions that are well reconstructed using CoRRECT. The $R_2^\ast$  corresponding to \emph{TV},  \emph{RED}, and \emph{DU} are obtained by the recent LEARN-BIO network~\cite{Xu.etal2022}. Note the excellent quantitative performance of CoRRECT for mGRE reconstruction and $R_2^\ast$ estimation.}
		\label{fig:simu_across_method}
	\end{figure*}
	
	\subsection{Experimental Setup}
	In this section, we present the implementation details and baselines for evaluating the performance of CoRRECT on mGRE image reconstruction and $R_2^\ast$ map estimation.
	
	\subsubsection{Baseline Methods for mGRE Reconstruction} 
	We  considered several well-known algorithms for image reconstruction, including  \emph{TV}~\cite{Rudin.etal1992},  \emph{U-Net}~\cite{Ronneberger.etal2015}, and \emph{RED}~\cite{Romano.etal2017}. We also investigated a \emph{deep unfodling (DU)} network identical to the image reconstruction module of CoRRECT. Inclusion of DU helps to illustrate improvements due to joint training. TV is a traditional optimization-based method that does not require training, while other methods are all DL methods with publicly available implementations. We trained the DL methods on motion-free data to handle  mGRE reconstruction. We used the same DnCNN~\cite{Zhang.etal2017} architecture used in our image reconstruction module as the AWGN denoiser for RED. The RED denoisers were trained for AWGN removal at four noise levels corresponding to noise variances $\sigma\in$\{1, 3, 5, 7\}. For each experiment with RED, we selected the denoiser achieving the highest SNR. DU shares the same setting as our image reconstruction module, except that it was trained on the motion-free data. We ran TV and RED both for 50 iterations. We fixed the step size $\gamma = 0.5$  for TV, RED, DU, and CoRRECT. We used \texttt{fminbound} in the \texttt{scipy.optimize} toolbox to identify the optimal regularization parameters $\tau$ for TV and RED at the inference time, and learned its value through training for DU and CoRRECT.

	\subsubsection{Baseline Methods for $R_2^\ast$ Estimation} 
	We applied the DL-based $R_2^\ast$ estimation method LEARN-BIO~\cite{Xu.etal2022} to the reconstructed mGRE images from baseline methods to compute the corresponding $R_2^\ast$ maps  as comparisons to the ones from our end-to-end training. LEARN-BIO used the same architecture as $\Esf_\varphibm$ of CoRRECT, except that it was not jointly trained with $\Rsf_\thetabm$.  We trained two LEARN-BIO networks, namely LEARN-BIO (clean) and LEARN-BIO (motion). LEARN-BIO (clean) was trained on motion artifact-free mGRE images and applied to ground-truth mGRE images to get ground-truth $R_2^\ast$ reference images for quantitative evaluation. LEARN-BIO (motion) was trained on motion-corrupted mGRE images (generated with the same motion simulation configuration introduced in Sec.~\ref{sec:data_simu}) to compute motion-corrected $R_2^\ast$ maps. We applied this network to all reconstruction methods to see their ability to produce $R_2^\ast$ maps. As an additional reference, we used the traditional voxel-wise NLLS approach to estimate $R_2^\ast$ maps from subsampled, noisy, and motion-corrupted mGRE images reconstructed using zero-filling (ZF), denoted as \textit{ZF+NLLS}. As described in Sec.~\ref{sec:biophysicalmodel}, NLLS is a standard iterative fitting method for computing  $R_2^\ast$ based on Eq.~\eqref{Eq:Model}, where in each iteration, the regression is conducted by combining the data from different echo times $t$ with their $F(t)$ values voxel by voxel. \sub{Prior to the NLLS fitting, a brain extraction tool, implemented in the Functional Magnetic Resonance Imaging of the Brain Library (FMRIB), was used to generate REMs to mask out both skull and background voxels in all mGRE data~\cite{Jenkinson.etal2005}, where the signal model defined in Eq.~\eqref{Eq:Model} does not apply.} NLLS was run over only the set of unmasked voxels. We used the same REMs in the loss function Eq.~\eqref{eq:loss-est} for training $R_2^\ast$ estimation module as well as in the baseline LEARN-BIO method. All the results of $R_2^\ast$ presented in this paper were also masked before visualization.
	
	\begin{figure*}[t]
		\centering 
		\includegraphics[width=\textwidth]{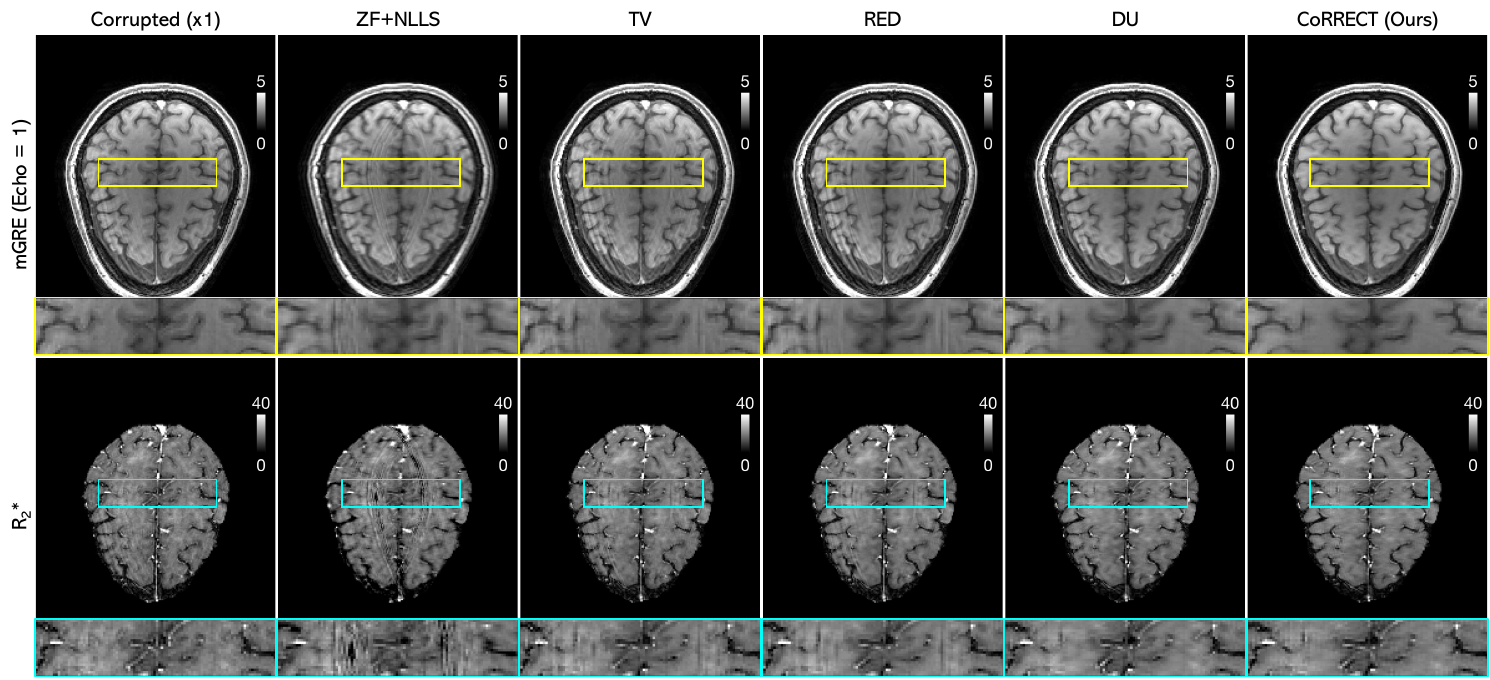}
		\caption{Visual evaluation of CoRRECT on experimentally collected data corrupted with real motion, sampled using acceleration factor $\times 4$. The mGRE image in the first column (denoted with $\times 1$) uses motion-corrupted but fully-sampled k-space data, while the ones in other columns use motion-corrupted and subsampled k-space data. Note the excellent performance of CoRRECT for producing high-quality mGRE and $R_2^\ast$ images. Note also the abilty of CoRRECT trained on synthetic motion to address artifacts due to real object motion.}
		\label{fig:real_across_method}
	\end{figure*}
	
	\subsubsection{Implementation Details and Evaluation Metrics}
	We used the $\ell_2$ loss for $ \mathcal{L}_\mathrm{rec}(\thetabm)$  and $\mathcal{L}_\mathrm{est}(\thetabm, \varphibm)$ with the weighting parameter $\lambda = 1$. We used the Adam~\cite{Kingma.Ba2015} optimizer to minimize the loss and set the initial learning rate to $1 \times 10^{-5}$. We performed all our experiments on a machine equipped with 8 GeForce RTX 2080 GPUs. For quantitative evaluation,  we used \emph{signal-to-noise ratio (SNR)}, measured in dB, and \emph{structural similarity index (SSIM)}, relative to the ground-truth. Since the ground-truth was not available for experimental data, we provided qualitative visual comparisons of different methods.
	
	\begin{figure*}[t]
		\centering 
		\includegraphics[width=\textwidth]{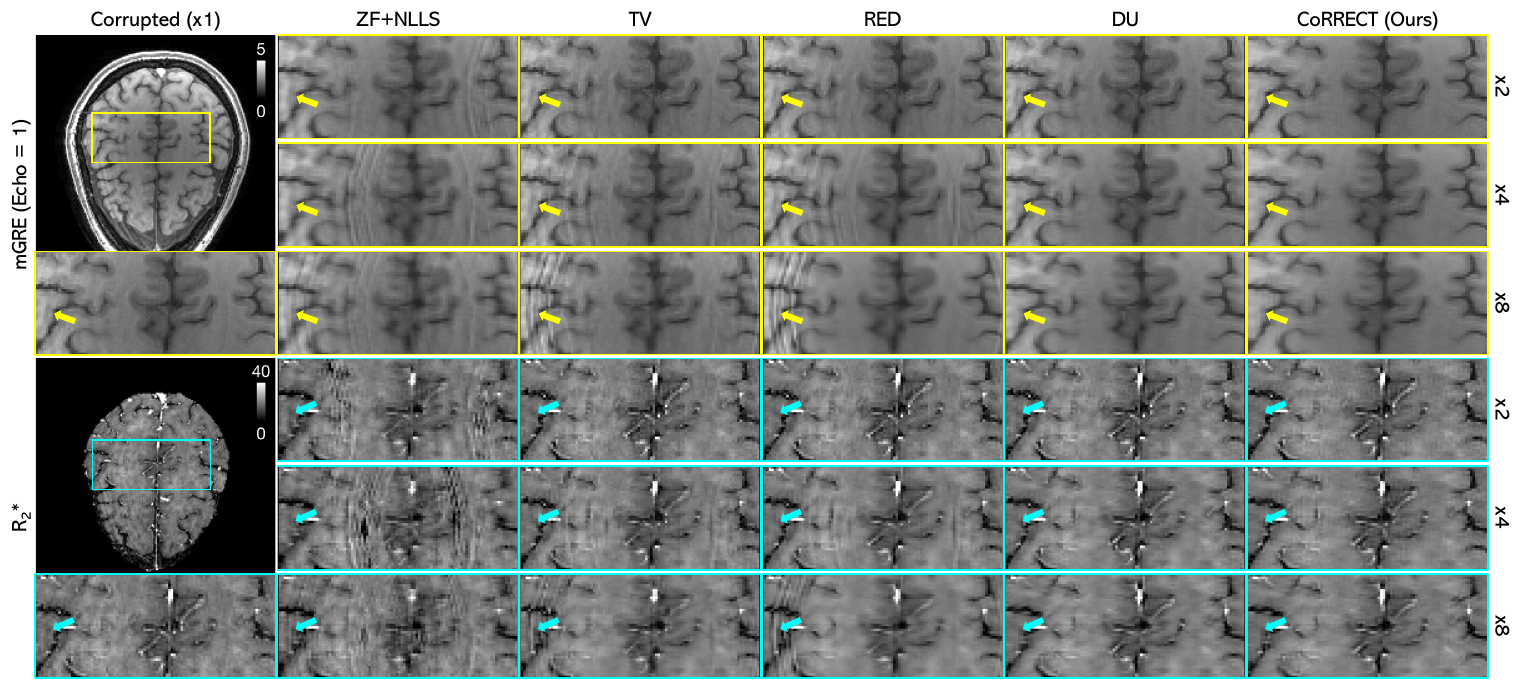}
		\caption{Visual evaluation of CoRRECT on experimentally-collected data corrupted with real motion, subsampled using acceleration rates $\{\times 2, \times 4, \times 8 \}$. Arrows in the zoomed-in plots highlight brain regions that are well reconstructed using CoRRECT. \emph{Corrupted ($\times 1$)} uses motion-corrupted but fully-sampled measurements, while \emph{ZF+NLLS}, \emph{TV}, \emph{RED}, \emph{DU} and \emph{CoRRECT} use motion-corrupted and subsampled measurements.  Note the improvements due to CoRRECT across different sampling rates.}
		\label{fig:real_248}
	\end{figure*}
	
	\subsection{Results on Simulated Data}
	\label{sec-simulated}
	We first tested the performance of CoRRECT on simulated data with synthetic motion corruptions. We followed the configuration in  Sec.~\ref{sec:data_simu} to add random motion to each data slice in our testing dataset to cover comprehensive motion levels. \sub{Table~\ref{tab:simu}  summarizes quantitative results of all evaluated methods at different acceleration rates, and Fig.~\ref{fig:statistic} visualizes the  statistics of the results with box plots.} \textcolor{xj}{As highlighted in Table~\ref{tab:simu} and Fig.~\ref{fig:statistic}, CoRRECT achieved the highest averaged SNR and SSIM values compared to other methods over all considered configurations.}
	
	Fig.~\ref{fig:simu_across_method} shows the performance of CoRRECT compared with different baseline methods on exemplar simulated data. The $1$st echo of a complex-valued mGRE image sequence is shown as its normalized magnitude, where the normalization was done by dividing by the mean of the intensity in the $1$st echo of the mGRE sequence. The result of ZF+NLLS showed that subsampling and motion can severely degrade the quality of mGRE images by causing a significant amount of blurring and aliasing artifacts, and consequently leads to suboptimal $R_2^\ast$ estimation.  Baseline methods TV and RED alleviated some of the artifacts in the corrupted image. However, due to their inability to capture the motion effects missed in the forward operator,  a considerable amount of  artifacts were still observed in mGRE reconstruction. Meanwhile,  due to the existence of unknown motion, the forward operator $\Abm$ that only models the subsampling was no longer accurate and consequently misled the reconstruction. DU further reduced the overall artifacts by using a CNN prior to compensate for artifacts through end-to-end training, but was still suboptimal, showing visible artifacts in mGRE reconstruction.   As for $R_2^\ast$  estimation, although a significant improvement over the NLLS fitting was observed by using motion-correction-enabled LEARN-BIO on artifact-contaminated mGRE images from those baseline methods, the estimation still suffered from inaccuracy in the regions indicated by blue arrows. Our proposed method, CoRRECT, managed to achieve the best performance compared to all evaluated baseline methods in terms of sharpness, contrast, artifact removal and accuracy, thanks to joint training of mGRE reconstruction and  $R_2^\ast$ estimation.

	\begin{figure*}[t]
		\centering 
		\includegraphics[width=\textwidth]{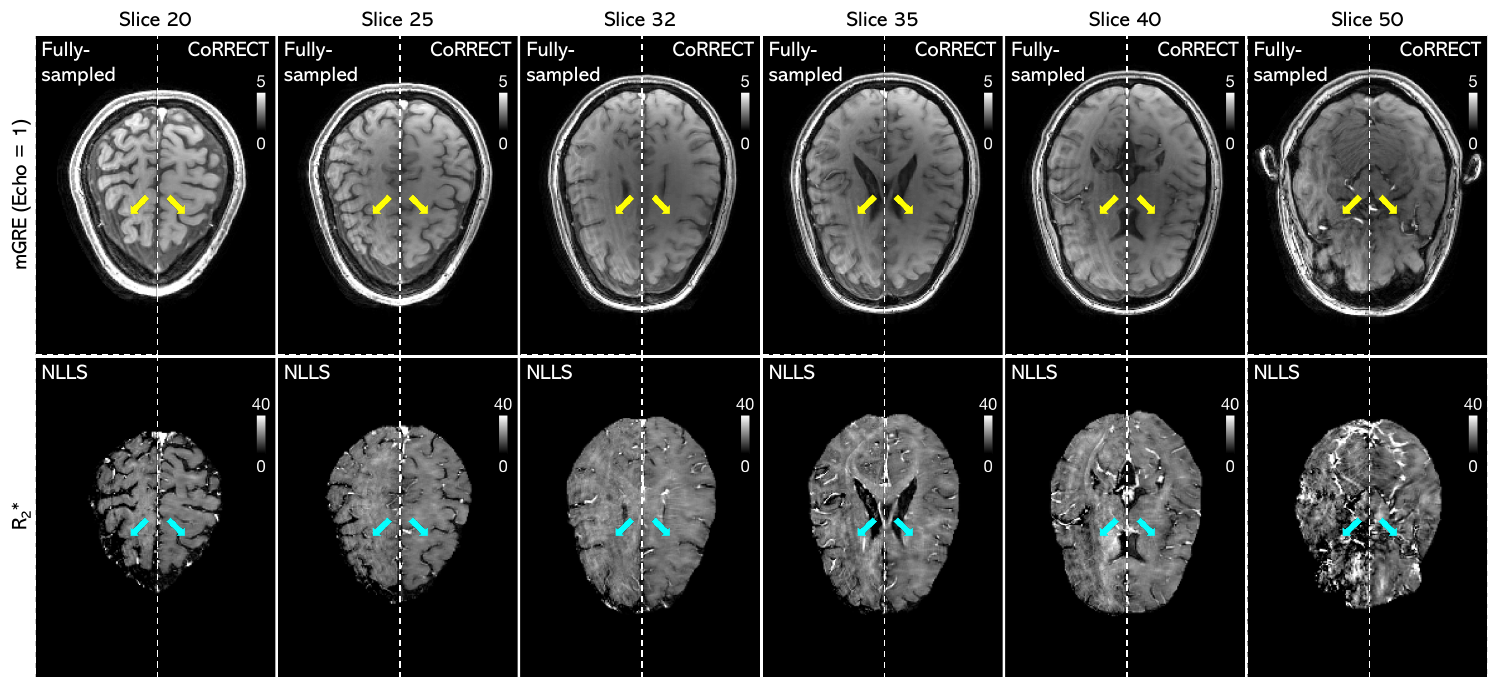}
		\caption{ Visual evaluation of CoRRECT on experimental data corrupted with real motion, sampled using acceleration rate $\times 4$. The first row shows several slices of reconstructed mGRE images from the whole brain volume of 72 slices, while the second row shows the corresponding estimated $R_2^\ast$ maps. In each column of the first row, the images to the left of the dashed line are the mGRE images reconstructed from the \emph{fully-sampled}, noisy, and motion-corrupted measurements, while the images to the right are the result of the CoRRECT reconstruction from \emph{subsampled}, noisy, and motion-corrupted measurements.  In each column of the second row, the $R_2^\ast$ maps to the left of the dashed line are estimated using NLLS on mGRE images in the first row, while those to the right are produced by CoRRECT. Arrows in the plots highlight brain regions that are well reconstructed using CoRRECT. Note how CoRRECT can remove artifacts across the whole brain volume.}
		\label{fig:real_across_slices}
	\end{figure*}
	
	\subsection{Results on Experimental Data }
	\label{sec-simulated-real}
	We further validated the performance of our network trained on simulated data using experimental data with real motion corruptions. 
	
	Fig.~\ref{fig:real_across_method} shows the performance of CoRRECT compared to different baseline methods on exemplar experimental data corrupted with real motion and subsampled with acceleration rate $\times 4$. Note that the corrupted mGRE image in the first column, denoted with acceleration rate $\times 1$,  corresponds to the  corrupted mGRE image of motion-affected but fully-sampled k-space data. The corresponding $R_2^\ast$,  which was estimated using LEARN-BIO (clean),  consequently  suffered from these motion corruptions as well. While such motion artifacts in this experimental data might not follow our simulation model, we did observe similar results to our synthetic experiments.  It can be seen that CoRRECT outperformed the evaluated baseline methods in both mGRE reconstruction and  $R_2^\ast$ estimation in terms of removing artifacts and maintaining sharpness. This shows CoRRECT is capable of handling real motion artifacts while still keeping detailed structural information. Fig.~\ref{fig:real_248} shows comprehensive results across different acceleration rates for the same data sample, where consistently outstanding and robust performance of CoRRECT was observed.
	
	Fig.~\ref{fig:real_across_slices} further demonstrates the performance of our method across different data slices in a whole brain volume, where each slice, in principle,  was corrupted with different and random motions during the scan. For each slice, we show the side-to-side  comparison between the results of CoRRECT on the subsampled (with acceleration rate $\times 4$), noisy and motion-corrupted measurements and the images obtained from fully-sampled, noisy, and motion-corrupted k-space measurements and their NLLS-estimated  $R_2^\ast$ maps. Note that CoRRECT using subsampled measurements successfully removed the motion artifacts visible in the fully-sampled images. The constant success of CoRRECT on different brain slices proved that our network can work on the whole spectrum of brain volume, highlighting the effectiveness and adaptability of our method.

	\begin{table*}[t]
		\centering
		{
			\footnotesize
			\begin{threeparttable}
				\caption{\textcolor{sub}{Quantitative comparison between separately-trained and jointly-trained models on simulated motion-corrupted data at different sampling rates.}}
				\label{tab:benefit_of_joint_training}
				\renewcommand\arraystretch{1}
				\setlength{\tabcolsep}{6pt}
				\begin{tabular}{ccccccc} 
					\toprule				
					\textit{\textcolor{sub}{Images}}       & \multicolumn{6}{c}{\textcolor{sub}{$mGRE$}} \\ 
					\cmidrule(rl){2-7}
					\textcolor{sub}{\textit{Metric}}  &\multicolumn{3}{c}{\textcolor{sub}{SNR (dB)}} & \multicolumn{3}{c}{\textcolor{sub}{SSIM}}\\
					\cmidrule(rl){2-7}
					
					\textcolor{sub}{\textit{Acceleration rate}}   & \textcolor{sub}{x2} & \textcolor{sub}{x4} & \textcolor{sub}{x8} & \textcolor{sub}{x2} & \textcolor{sub}{x4} & \textcolor{sub}{x8} \\
					\cmidrule(rl){2-4}\cmidrule(rl){5-7}
					
					\textcolor{sub}{Separately-trained} & \textcolor{sub}{22.08} &  \textcolor{sub}{\textbf{20.69}} &\textcolor{sub}{\textbf{19.28}}  &\textcolor{sub}{0.93} &\textcolor{sub}{0.91}  &\textcolor{sub}{0.88}\\
					\textcolor{sub}{Jointly-trained (Ours)} &\textcolor{sub}{\textbf{22.12}} & \textcolor{sub}{{20.66}} & \textcolor{sub}{{19.25}} & \textcolor{sub}{\textbf{0.93}} & \textcolor{sub}{\textbf{0.91}} & \textcolor{sub}{\textbf{0.90}} \\
					\bottomrule
				\end{tabular}
				
				\begin{tabular}{ccccccc} 
					\toprule				
					\textit{\textcolor{sub}{Images}}       & \multicolumn{6}{c}{\textcolor{sub}{$R_2^\ast$}} \\ 
					\cmidrule(rl){2-7}
					\textcolor{sub}{\textit{Metric}}  &\multicolumn{3}{c}{\textcolor{sub}{SNR (dB)}} & \multicolumn{3}{c}{\textcolor{sub}{SSIM}}\\
					\cmidrule(rl){2-7}
					
					\textcolor{sub}{\textit{Acceleration rate}}   & \textcolor{sub}{x2} & \textcolor{sub}{x4} & \textcolor{sub}{x8} & \textcolor{sub}{x2} & \textcolor{sub}{x4} & \textcolor{sub}{x8} \\
					\cmidrule(rl){2-4}\cmidrule(rl){5-7}
					
					\textcolor{sub}{Separately-trained} & \textcolor{sub}{11.94} &  \textcolor{sub}{11.51} &\textcolor{sub}{10.95}  &\textcolor{sub}{0.91} &\textcolor{sub}{0.90}  &\textcolor{sub}{0.88}\\
					\textcolor{sub}{Jointly-trained (Ours)} &\textcolor{sub}{\textbf{12.99}} & \textcolor{sub}{\textbf{12.33}} & \textcolor{sub}{\textbf{11.60}} & \textcolor{sub}{\textbf{0.92}} & \textcolor{sub}{\textbf{0.90}} & \textcolor{sub}{\textbf{0.89}} \\
					\bottomrule
				\end{tabular}
				
			\end{threeparttable}
		}
	\end{table*}

\begin{figure*}
		\centering
		\includegraphics[width=.6\textwidth]{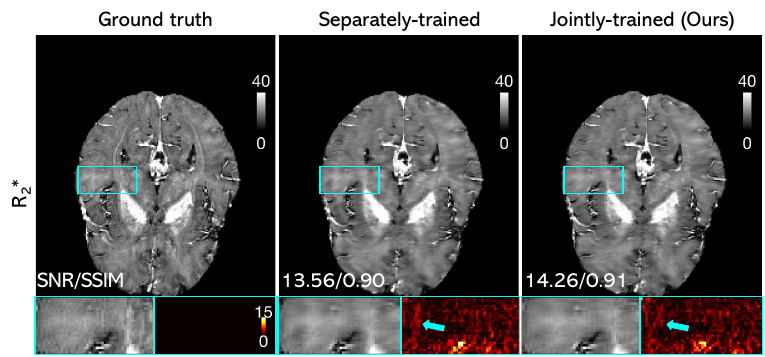}
		\caption{\sub{Visual comparison between separately-trained and jointly-trained models on simulated motion-corrupted data at sampling rates $\times 4$. The bottom-left corner of each image provides the SNR and SSIM values with respect to the ground-truth.  Arrows in the zoomed-in plots highlight brain regions that are well reconstructed using jointly-trained CoRRECT. Note the excellent quantitative performance of our jointly-trained model for end-to-end $R_2^\ast$ estimation.}}
		\label{fig:benefit_of_joint_training}
	\end{figure*}

	\sub{\subsection{Benefits of Joint Training}
		To demonstrate the benefits of joint training used in CoRRECT, we compared the $R_2^\ast$ estimation performance of (a) the combined but separately-trained MRI reconstruction sub-module and the qMRI estimation sub-module of CoRRECT  with (b) our proposed end-to-end and jointly-trained CoRRECT model. Both the separately-trained and jointly-trained models were trained on our simulated motion-corrupted training dataset and tested on our simulated motion-corrupted testing dataset. Table~\ref{tab:benefit_of_joint_training} shows the comprehensive quantitative performance of both models averaged on  our testing dataset at different sampling rates and Fig.~\ref{fig:benefit_of_joint_training} shows the visual performance of an exemplar $R_2^\ast$ estimation of our testing data. Note that while both the separately-trained and jointly-trained models achieved good mGRE reconstruction, the consistently superior quantitative performance of $R_2^\ast$ estimation in the jointly trained model, as shown in Table~\ref{tab:benefit_of_joint_training}, along with the improved visual performance in Fig.~\ref{fig:benefit_of_joint_training}, underscores the advantages of CoRRECT's joint training approach. } 
	
	

	\section{Discussion and Conclusion}
	\label{sec:conclusion}
	This work introduces CoRRECT as a framework for estimating $R_2^\ast$ maps from subsampled, noisy, and artifact-corrupted mGRE k-space measurements. Unlike existing DL techniques for MRI that separate quantitative parameter estimation from image reconstruction, CoRRECT addresses both tasks by integrating three core components: (a) an end-to-end model-based neural network, (b) a training scheme accounting for motion-artifacts, and (c) a loss function for training without ground-truth $R_2^*$ maps. Our extensive validation corroborate the potential of DMBAs integrating multiple models to produce high-quality images from noisy, subsampled, and artifact-corrupted measurements.

	\section{Acknowledgment}
	\label{sec:acknowledgment}
	Research reported in this publication was supported by the NSF CAREER award CCF-2043134 and the National Institutes of Health (NIH) awards R01AG054513,  RF1AG077658, and RF1AG082030.


	
	\bibliography{sn-bibliography}
\end{document}